\documentclass[pra,showpacs,nofootinbib]{revtex4}
\usepackage{hyperref}
\usepackage{amsmath}
\usepackage{amssymb}
\usepackage{amsthm}
\usepackage{graphics}
\usepackage{graphicx}
\usepackage{color}
\usepackage[usenames,dvipsnames,svgnames,table]{xcolor}
\usepackage{enumitem}

\long\def\ca#1\cb{}

\def\bra#1{\langle#1\vert}

\def\inpd#1#2{\langle#1\vert#2\rangle }
\def\ket#1{\vert#1\rangle }
\def\proj#1#2{\vert#1\rangle_#2\langle#1\vert}

\def\Tr#1{\textrm{Tr}(#1)}
\def\AC{{\cal A}}
\def\BC{{\cal B}}
\def\CC{{\cal C}}

\def\EC{{\cal E}}

\def\KC{{\cal K}}

\def\SC{{\cal S}}

\newtheorem{thm1}{Theorem}

\newtheorem{thm4}[thm1]{Theorem}

\newtheorem{cor1}[thm1]{Corollary}

\begin{document}
\title{On the structure of LOCC: finite vs. infinite rounds}
\author{Scott M. Cohen}
\email{cohensm52@gmail.com}
\affiliation{Department of Physics, Portland State University, Portland OR 97201}

\begin{abstract}
Every measurement that can be implemented by local quantum operations and classical communication (LOCC) using an infinite number of rounds is the limit of a sequence of measurements each of which requires only a finite number of rounds. This rather obvious and well-known fact is nonetheless of interest as it shows that these infinite-round measurements can be approximated arbitrarily closely simply by using more and more rounds of communication. Here we demonstrate the perhaps less obvious result that (at least) for bipartite systems, the reverse relationship also holds. Specifically, we show that every finite-round bipartite LOCC measurement is the limit of a continuous sequence of LOCC measurements, where each measurement in that sequence can be implemented by LOCC, but only with the use of an infinite number of rounds. Thus, the set of LOCC measurements that require an infinite number of rounds is dense in the entirety of LOCC, as is the set of finite-round LOCC measurements. This means there exist measurements that can only be implemented by LOCC by using an infinite number of rounds, but can nonetheless be approximated closely by using one round of communication, and actually in some cases, no communication is needed at all. These results follow from a new necessary condition for finite-round LOCC, which is extremely simple to check, is very easy to prove, and which can be violated by utilizing an infinite number of rounds.
\end{abstract}

\date{\today}

\maketitle
\section{Introduction}\label{sec1}
Measurements play a critical role in all areas of experimental science. It is therefore necessary to understand measurements from a theoretical point of view, to answer questions as to what measurements are possible, how their results can be interpreted, and what precisely they can tell us about the system being measured. Such questions have taken on renewed importance since the discovery of quantum mechanics, one reason being the so-called ``measurement problem", a problem closely related to early foundational questions raised by Schr\"{o}dinger's cat \cite{Schrodinger1}, which has been much debated ever since \cite{RBGMeasProb,Schlosshauer,JoosZeh,Wigner}. With the recent explosion of interest in quantum information and quantum computation, new questions have arisen, including in the general area we will be concerned with here, that of measurements performed by multiple parties on spatially separated quantum systems.

If the parties lack the ability to communicate quantum information, then they are restricted to performing quantum operations on their individual local subsystems and then communicating classical information about the results of their actions to the other parties. The study of these local operations and classical communication (LOCC) protocols has proven to be extremely challenging, but they are widely applicable in the context of quantum information theory, with examples of such applications ranging from distributed quantum computing \cite{CiracDistComp}, one-way quantum computing \cite{RaussendorfBriegel}, entanglement distillation \cite{BennettPurify} and manipulation \cite{Nielsen}, to local distinguishability of quantum states \cite{Walgate}, local cloning \cite{Anselmi,ourLocalCloning}, and various quantum cryptographic protocols, such as secret sharing \cite{HillerySecret}. As a result, efforts to understand these processes have been extensive.

Every LOCC protocol implements a separable measurement (SEP) \cite{Rains}, and much can be learned by studying the latter set. It is known, however, that there exist separable measurements that are not LOCC \cite{Bennett9}. Therefore, the study of SEP for the purpose of gaining insights about LOCC will best be accompanied by an understanding of exactly how these two sets of measurements differ. Much has been learned about this difference in recent years, including several results showing a gap between what can be accomplished by SEP as opposed to by LOCC for specific tasks such as quantum state discrimination \cite{Bennett9,IBM_CMP,NisetCerf,myLDPE,MaWehnerWinters,KKB,ChildsLeung,ChitHsieh1,ChitDuanHsieh,FuLeungManciska,ChitHsieh2} and the transformation of entangled quantum states \cite{FortescueLo,Chitambar,ChitCuiLoPRL,ChitCuiLoPRA,WinterLeung}. In the latter case, it has been shown that this gap can be sizable \cite{ChitCuiLoPRL,ChitCuiLoPRA}

A number of studies have considered the significance of the number of rounds of communication used by the parties. In the bipartite case, it has been shown that for the task of transforming a system from one pure state to another, multiple rounds is not better than a single one \cite{LoPopescu}. In contrast, there are mixed-state purification scenarios which require at least two rounds of communication \cite{BennettPurify}. The question of whether or not it can be helpful to use an infinite number of rounds has also been studied. In \cite{KKB}, it was shown that this is not helpful for perfect state discrimination of complete product bases, but certain random distillation tasks \cite{FortescueLo} were used to demonstrate circumstances where infinite rounds are required \cite{Chitambar}. The latter example involved three parties, but a bipartite case has also been given \cite{WinterLeung} using similar ideas.

In the present work, we continue this study of infinite-round LOCC with a focus on the difference between finite-round protocols and those that require infinite rounds. In Section~\ref{sec2}, we present a new necessary condition that a measurement can be implemented with finite-round LOCC. In Section~\ref{sec3}, we give a specific example of a two-qubit measurement that violates this condition, and then we construct an infinite-round LOCC protocol that implements this measurement exactly. This necessary condition therefore provides the first completely general approach (by which we mean an approach applying to every LOCC measurement, as opposed to a subset such as measurements useful for state discrimination \cite{KKB,ChitHsieh2}) to distinguishing between measurements that can be implemented by finite-round LOCC and those that can only be implemented using an infinite number of rounds. In Section~\ref{sec4} we demonstrate that this two-qubit example can be readily generalized, constructing a general class of measurements on any bipartite system which can be implemented by using an infinite number of rounds, but not with any finite-round protocol. 

Using these ideas, we then show in Section \ref{sec5} how any finite-round LOCC protocol can be extended to infinite rounds in a way such that (i) the measurement implemented by the resulting protocol violates our necessary condition for finite-round LOCC, but (ii) in the limit of a particular small parameter approaching zero, the original finite-round protocol is recovered. These results lead directly to the main results of this paper, in which we use the following definitions: If a measurement, ${\bf M}$, can be implemented in a finite number of rounds, we say that ${\bf M}\in$ LOCC$_\mathbb{N}$. If it can be implemented by LOCC but only by using an infinite number of rounds, then ${\bf M}\in$ LOCC$\backslash$LOCC$_\mathbb{N}=:$ LOCC$_\infty$ (here, we define LOCC as including all measurements implemented by finite- or infinite-round protocols; this corresponds to what is denoted as $\overline{\textrm{LOCC}}$ in \cite{WinterLeung}). We also denote the boundary of set $X$ as $\partial X$. Our main results can then be stated as the following theorem and its corollary.
\begin{thm4}\label{thm4}
Every measurement ${\bf M}_0\in$ LOCC$_{\mathbb{N}}$ is a limit point of the set of measurements LOCC$_\infty$. That is, for every ${\bf M}_0\in$ LOCC$_{\mathbb{N}}$ there exists a continuous sequence of measurements, ${\bf M}_\epsilon\in$ LOCC$_\infty,~\epsilon>0$, such that ${\bf M}_0=\lim_{\epsilon\rightarrow0}{\bf M}_\epsilon$.
\end{thm4}
\noindent  In order to have a reasonable notion of limit point, as used in this theorem, we introduce a distance measure between measurements in the following subsection. Using this theorem along with the obvious and well-known fact that every measurement in LOCC$_\infty$ can be arbitrarily closely approximated by measurements in LOCC$_{\mathbb{N}}$, we have this corollary.
\begin{cor1}\label{cor1}
The following related results are direct consequences of Theorem~\ref{thm4}.
\begin{enumerate}
  \item Both LOCC$_{\mathbb{N}}$ and LOCC$_\infty$ are dense in LOCC;
  \item The boundaries of these subsets of LOCC satisfy $\partial$LOCC$_{\mathbb{N}}=$LOCC, and LOCC $\subset\partial$LOCC$_\infty$;
  \item The closure of LOCC$_{\mathbb{N}}$ is LOCC, and LOCC is strictly contained within the closure of LOCC$_\infty$; and
  \item LOCC$_{\mathbb{N}}$ and LOCC$_\infty$ each has empty interior.
\end{enumerate}
\end{cor1}
\noindent The strict inclusion in the second and third items of this corollary follow directly from a result of \cite{ChitCuiLoPRL}, where it was shown that there exist sequences of measurements in LOCC$_\infty$ whose limit is not in LOCC. Note also that the last item about the interior of LOCC$_{\mathbb{N}}$ appears to contradict a result of \cite{WinterLeung}; see the discussion in Section~\ref{sec6} for an explanation.

\subsection{Definitions and terminology}\label{subsec11}
Before proceeding, let us review basic ideas on quantum measurements and introduce terminology. A quantum channel is a completely positive, trace-preserving map $\EC$ from one operator space to another. It can be described in terms of a set of Kraus operators $\{K_j\}$ \cite{Kraus} as
\begin{align}\label{eqn02}
\rho^\prime=\EC(\rho)&=\sum_jK_j\rho K_j^\dag,\\
\sum_jK_j^\dag K_j&=I,\label{eqn03}
\end{align}
where $I$ is the identity operator on the full input space and $\rho$ ($\rho^\prime$) are operators acting on that of the input (output). (Note that the set of Kraus operators describing a given channel is not unique.) By ``quantum measurement", on the other hand, we will mean a quantum channel that also outputs the classical index $j$,
\begin{align}\label{eqn04}
\rho\rightarrow\rho_j^\prime=K_j\rho K_j^\dag/p_j,~p_j=\Tr{K_j^\dag K_j\rho},
\end{align}
which can also be represented as
\begin{align}\label{eqn06}
\EC(\rho)&=\sum_jK_j\rho K_j^\dag\otimes\ket{j}\bra{j}.
\end{align}
Finally, a POVM is a measurement where the output is discarded and only the probabilities, $p_j$, are retained. Each positive (semidefinite) operator $K_j^\dag K_j$ is referred to as a POVM element.

We now introduce a distance measure between measurements. Consider two measurements, ${\bf M}_1,{\bf M}_2$, each defined in terms of its set of distinct Kraus operators, $\{K_{1j}\}$ and $\{K_{2j}\}$, respectively. We say that two Kraus operators are distinct if they are not proportional to each other (if two Kraus operators in a given measurement are proportional, say $c_1K$ and $c_2K$, we combine these into a single operator, $K^\prime=\sqrt{|c_1|^2+|c_2|^2}K$, in our definition of the measurement). If the number of distinct Kraus operators differs between ${\bf M}_1$ and ${\bf M}_2$, we pad the smaller set with zero-operators so that the two sets will have the same number (which may be infinite). Then our distance measure is
\begin{align}\label{eqn05}
d\left({\bf M}_1,{\bf M}_2\right)=\sup_\rho\inf_f\sum_j\left\|K_{1f_j}\rho K_{1f_j}^\dag-K_{2j}\rho K_{2j}\right\|_2,
\end{align}
\noindent where the supremum is taken over all density operators $\rho\ge0$, $\Tr{\rho}=1$, the infimum is taken over all permutations $f$ (that is, $f$ is a bijection mapping a subset of the natural numbers to itself), and $\|X\|_2=\max\{\sqrt{\lambda}|\lambda \textrm{ is an eigenvalue of } X^\dag X\}$ is the spectral norm of operator $X$. One can readily show that \eqref{eqn05} obeys the required properties of a distance measure: it is non-negative, it vanishes if and only if ${\bf M}_1={\bf M}_2$, it is symmetric, and it satisfies the triangle inequality. In addition, using \eqref{eqn03}, one can show that $0\le d\left({\bf M}_1,{\bf M}_2\right)\le 2$. We note also that it is not difficult to show the important property, used below, that $d\left({\bf M}_1,{\bf M}_2\right)$ is continuous; in particular, it continuously approaches zero as ${\bf M}_1\rightarrow {\bf M}_2$.

An LOCC measurement is a quantum measurement that is implemented by an LOCC protocol. Various different LOCC protocols may implement the same set of Kraus operators, but we will take the view that the resulting measurements are the same. It is well-known that each LOCC measurement is separable, by which is meant the corresponding Kraus operators are all product operators, of the form $K_j=K_j^{(1)}\otimes K_j^{(2)}\otimes\cdots\otimes K_j^{(P)}$ when there are $P$ parties involved (for a quantum channel, it just means there exists at least one set of product Kraus operators for the given channel). An LOCC protocol can be represented by a tree, where the children of any given node represent the outcomes of a measurement by one of the parties. For example, if the parties have collectively implemented Kraus operator $\hat K=\hat K^{(1)}\otimes\hat K^{(2)}\otimes\cdots\otimes\hat  K^{(P)}$ up to a given node, and then party $1$ measures with outcomes $K_i^{(1)}$, the resulting cumulative action becomes $\left(K_i^{(1)}\hat K^{(1)}\right)\otimes\hat  K^{(2)}\otimes\cdots\otimes\hat  K^{(P)}$, the local Kraus operators all remaining the same apart from that for party $1$, which changes to $K_i^{(1)}\hat K^{(1)}$ for each of the outcomes $i$. We have here introduced a notation that will be followed throughout the remainder of this paper, that Kraus operators representing the accumulated action up to a given point will have a hat ($\hat~$), while those representing an individual measurement (those operators transforming from a parent to its children in the tree) will not. Similarly, a measurement corresponding to cumulative Kraus operators by a single party will be denoted by $\hat M_j$, whereas an individual measurement by a single party will not have a hat. Overall measurements, which for our purposes will generally involve multiple parties, are denoted with bold-face, such as ${\bf M}_0$.  For each (cumulative) Kraus operator $\hat K_j$, we will denote the corresponding POVM element in calligraphic script, $\hat\KC_j=\hat K_j^\dag\hat K_j$ and $\hat\KC_j^{(\alpha)}=\hat K_j^{(\alpha)\dag}\hat K_j^{(\alpha)}$ for each party $\alpha$. We can also make use of the polar decomposition of an operator to write the Kraus operator in terms of the POVM element. Specifically, there exists isometry $U_j^{(\alpha)}$ such that
\begin{align}\label{eqn09}
\hat K_j^{(\alpha)} &= U_j^{(\alpha)}\sqrt{\hat\KC_j^{(\alpha)}},
\end{align}
and we take the unique positive-semidefinite square root of $\hat\KC_j^{(\alpha)}$.

A given measurement can be described in terms of its set of Kraus operators. Even though as we've defined our terms above, a measurement is distinct from a POVM, every measurement nonetheless has a set of POVM elements that describe it. Although for a complete description of the measurement, one must know the Kraus operators, we have nonetheless seen elsewhere \cite{mySEPvsLOCC,myMany,myExtViolate1} that a description in terms of POVM elements can be very fruitful, even when we are concerned with the more general notion of ``measurement" used here. In fact, we have shown \cite{mySEPvsLOCC,myMany} that in order to determine whether or not a given measurement can be implemented by LOCC in a finite number of rounds, it suffices to consider the POVM elements associated with that measurement. There, the focus is on the set of local POVM elements in the given measurement that are not proportional to each other, for each party, and we refer to two such elements as ``distinct" if they are not proportional.

An LOCC tree consists of nodes branching to a set of child nodes. The nodes may be labeled by the local Kraus operator corresponding to the outcome of the individual measurement represented by that node, or by the local Kraus operator representing the cumulative action of that party up to that point in the protocol, or by the corresponding cumulative POVM element. Let us choose the latter, and let us also refer to a node as an $\alpha$ node if it corresponds to an outcome of a measurement by party $\alpha$. Then, any set of sibling $\alpha$ nodes (nodes that all share the same parent), say $\{\hat\KC_{ij}^{(\alpha)}\}$, are descendant from their closest ancestor $\alpha$ node, $\hat\KC_i^{(\alpha)}$. As discussed in \cite{mySEPvsLOCC}, these operators satisfy the following relation,
\begin{align}\label{eqn00}
	\hat\KC_i^{(\alpha)}=\sum_{j\in\textrm{siblings}}\hat\KC_{ij}^{(\alpha)}.
\end{align}
In addition, any tree satisfying this relation at every node and having the identity operator at the root nodes for all parties is a valid LOCC tree \cite{mySEPvsLOCC}. These ideas will play an important role in what follows.

We now turn to our necessary condition for finite-round LOCC.

\section{A necessary condition for finite-round LOCC}\label{sec2}
Here, we give a necessary condition for finite-round LOCC. This condition is applicable quite generally, including for any number of parties. There is one minor restriction, however. For any given measurement, we assume there are at least two distinct POVM elements. If this is not the case, all Kraus operators are proportional to product isometries, and it is clear that such measurements can be done using LOCC and needing no more than one round of communication; we omit such measurements from consideration in our necessary condition for finite-round LOCC (they violate it, but our proof does not apply to them). The proof of our necessary condition for finite-round LOCC is then quite simple and follows almost immediately from the evident observation that in any such finite-round protocol, someone makes a last measurement that has at least two outcomes whose POVM elements are distinct.

Consider a separable measurement with POVM elements $\hat\KC_j=\hat\KC_j^{(1)}\otimes\cdots\otimes\hat\KC_j^{(P)}$ and assume there are at least two such operators that are distinct. Each local operator $\hat\KC_j^{(\alpha)}$ generates a ray in the convex cone of positive operators acting on that party's Hilbert space, this ray being defined as $\{\lambda\hat\KC_j^{(\alpha)}\vert\lambda\ge0\}$. Let us define the ray deficit as follows. For each party $\alpha$, identify all distinct rays generated by the set of operators $\{\hat\KC_j^{(\alpha)}\}$. If there exists $i\ne j$ such that $\hat\KC_i^{(\alpha)}\propto\hat\KC_j^{(\alpha)}$, then these two operators correspond to the same ray $r_\alpha$, and we then say that this ray has multiplicity $m_{r_\alpha}>1$. More generally, $m_{r_\alpha}$ is given by the number of distinct $\hat\KC_j$ whose local parts $\hat\KC_j^{(\alpha)}$ all lie on the same ray $r_\alpha$. Then, we define the ray deficit as
\begin{align}\label{eqn21}
\Delta=\sum_{\alpha=1}^P\sum_{r_\alpha}\left(m_{r_\alpha}-1\right).
\end{align}
Note that if the number of distinct operators $\hat\KC_j$ defining the separable measurement is finite, say $N$, then
\begin{align}\label{eqn22}
\Delta=PN-\sum_{\alpha=1}^PR_\alpha,
\end{align}
where $R_\alpha=\sum_{r_\alpha}(1)\le N$ is the total number of distinct rays generated by operators $\hat\KC_j^{(\alpha)}$. On the other hand, the quantity $\Delta$ can easily be finite and well-defined even in the limit $N\rightarrow\infty$ (and/or $P\rightarrow\infty$).

The minimum possible value of $\Delta$ for any separable measurement occurs when every ray corresponds to one, and only one, operator $\hat\KC_j^{(\alpha)}$. Under these conditions, $m_{r_\alpha}=1$ for all $r_\alpha$, and $\Delta=0$. For finite-round LOCC protocols, however, we have the following theorem.
\begin{thm1}\label{thm1}
For any finite-round LOCC protocol involving $P$ parties implementing a measurement corresponding to the set of distinct product POVM elements $\SC=\{\hat\KC_j\}$, where there are at least two elements in $\SC$, it must be that $\Delta\ge P-1$.
\end{thm1}
\proof Since there are at least two distinct operators in $\SC$, there must be at least one point in the protocol at which some party makes a last non-trivial measurement, where by `non-trivial', we mean it has at least two distinct POVM elements (any number of the parties may follow this measurement with an isometry). If, for example, it is party $\alpha$ that makes this last measurement, then the two outcomes of that measurement will be $\hat\KC_i^{(\alpha)}\not\propto\hat\KC_j^{(\alpha)}$. On the other hand, since one party's measurement does not change what the other parties have accomplished to that point, and since subsequent isometries by the parties do not alter their local POVM elements, these two final outcomes of the protocol, $\hat\KC_i,\hat\KC_j$, share the same local parts for all the other parties, $\hat\KC_i^{(\beta)}\propto\hat\KC_j^{(\beta)}$ for all $\beta\ne\alpha$. As a consequence, each of these latter $P-1$ operators correspond to rays $r_\beta$ having multiplicity $m_{r_\beta}\ge2$ and recalling the definition of $\Delta$ in \eqref{eqn21}, the theorem follows immediately.~\hspace{\stretch{1}}$\blacksquare$

In the next section, we will demonstrate that for bipartite systems, the bound in Theorem~\ref{thm1} can be violated, achieving $\Delta=0$, when using an infinite number of rounds of communication. This quantity $\Delta$ therefore provides a distinction between finite- and infinite-round LOCC. We also show in the appendix that for finite-round LOCC, $\Delta=1$ is achievable for $P=2$ and any number of distinct operators $\hat\KC_j$ (except when $N=1$, the case of all Kraus operators being isometries).

\section{A separable measurement requiring infinite rounds in LOCC}\label{sec3}
Let us here give an explicit example involving four Kraus operators acting on two qubits, which are a unique product representation of a separable quantum channel, and which can be implemented by LOCC with an infinite number of rounds, but not with any finite number of rounds. There are two parties $A$ and $B$ (for Alice and Bob), and the Kraus operators are $\hat A_j\otimes\hat B_j$ (these will be final outcomes of the LOCC protocol, so represent the cumulative action of the parties up to a given leaf), with
\begin{align}\label{eqn31}
\left\{\hat A_1,\hat A_2,\hat A_3,\hat A_4\right\}&=\left\{I_A,~\sqrt{1-q\,}[0]_A,~\sqrt{q\,}[0]_A+[1]_A,\,\sqrt{1-q\,}\ket{0}_A\bra{1}\right\},\nonumber\\
\left\{\hat B_1,\hat B_2,\hat B_3,\hat B_4\right\}&=\left\{\sqrt{1-\epsilon\,}[0]_B,~\sqrt{\epsilon\,}[0]_B+[1]_B,\sqrt{1-\epsilon\,}\ket{0}_B\bra{1},~\sqrt{\epsilon\,}I_B\right\},
\end{align}
where $0<\epsilon,q<1$ and, for example, $[0]_A=\proj{0}{A}$. The corresponding POVM elements are $\left(\hat \AC_j=\hat A_j^\dag\hat A_j,~\hat \BC_j=\hat B_j^\dag\hat B_j\right)$
\begin{align}\label{eqn32}
\left\{\hat \AC_1,\hat \AC_2,\hat \AC_3,\hat \AC_4\right\}&=\left\{I_A,~(1-q)[0]_A,~q[0]_A+[1]_A,~(1-q)[1]_A\right\},\nonumber\\
\left\{\hat \BC_1,\hat \BC_2,\hat \BC_3,\hat \BC_4\right\}&=\left\{(1-\epsilon)[0]_B,~\epsilon[0]_B+[1]_B,(1-\epsilon)[1]_B,~\epsilon I_B\right\}.
\end{align}
Note that $\sum_j\hat \AC_j\otimes\hat \BC_j=\left(1-q\epsilon\right)I_A\otimes I_B$, and therefore,
\begin{align}\label{eqn35}
{\bf M}_\epsilon=\left\{\frac{1}{\sqrt{1-q\epsilon}}\hat A_j\otimes\hat B_j\right\}_{j=1}^4
\end{align}
is a complete separable measurement. As no two of the $\hat\AC_j$ are proportional and no two of the $\hat\BC_j$ are proportional, this measurement satisfies $\Delta=0$, so by Theorem~\ref{thm1} it cannot be implemented by finite-round LOCC. Notice how simple this measurement is, involving only four operators acting on the smallest bipartite system of two qubits. In fact, as \eqref{eqn32} involves only diagonal matrices, it has the appearance of a classical measurement. Certainly, one can readily devise a different measurement that, after coarse-graining, reproduces the probability distribution of this one, and which can be implemented in few rounds. However, if the state of the system after the measurement is important, then it is not possible to reproduce the results that would be achieved by this measurement with any finite-round LOCC protocol. As can be seen from the results of \cite{myUniqueProdRep}, the Kraus operators of \eqref{eqn31} are the unique product representation for the corresponding channel (note that this would not be the case if we replace $\hat A_4\rightarrow\sqrt{1-q\,}[1]_A$ and $\hat B_3\rightarrow\sqrt{1-\epsilon\,}[1]_B$; this is the only reason these operators have been chosen non-diagonal, as all other conclusions remain valid with this replacement). It is therefore also not possible to implement this quantum channel by any finite-round LOCC, but we will now show how it can be implemented exactly with an infinite number of rounds.

The protocol involves four different local measurements repeated over and over, each of which has two outcomes. These four measurements are given by the following pairs of Kraus operators:
\begin{align}\label{eqn33}
\hat M_1&=\left\{\hat B_1,\hat B_2\right\},\nonumber\\
\hat M_2&=\left\{\hat A_2,\hat A_3\right\},\nonumber\\
\hat M_3&=\left\{\hat B_3,\sqrt{\epsilon\,}\hat B_2^{-1}\right\},\nonumber\\
\hat M_4&=\left\{\hat A_4,\sqrt{q\,}\hat A_3^{-1}\right\}.
\end{align}
The first outcome of each of these measurements terminates the protocol, while the second outcome will always be followed by another measurement. Bob starts with $\hat M_1$, and if he gets the second outcome $\hat B_2$, Alice measures with $\hat M_2$. If she gets the second outcome, Bob measures $\hat M_3$, and if he gets the second outcome, Alice does $\hat M_4$. If she gets the second outcome, their cumulative action to this point is the Kraus operator $\sqrt{q\epsilon\,}I_A\otimes I_B$, and so Bob can just start over with $\hat M_1$, the two of them continuing on ad infinitum. By comparing these to \eqref{eqn31}, one sees that the first outcome of each of these measurements is a ``correct" outcome (up to multiplicative factors) and so can be taken to terminate the protocol, whereas the second outcome is an error that needs to be corrected. For example, after the first outcome of Alice's measurement of $\hat M_2$, they have implemented the desired $\hat A_2\otimes \hat B_2$, but the second outcome leaves them with $\hat A_3\otimes \hat B_2$, which is an error since it is not one of the desired outcomes. The longer they continue, the smaller is the error, decreasing by a factor of $q\epsilon$ with every cycle through the four measurements $\hat M_j$, and vanishing in the limit of infinite rounds. Therefore in this limit, the protocol just described implements the measurement given in \eqref{eqn31}, with an overall weight factor (to the POVM elements $\hat\AC_j\otimes\hat\BC_j$) of $1+q\epsilon+(q\epsilon)^2+\cdots=\left(1-q\epsilon\right)^{-1}$.

Consider the limit $\epsilon\rightarrow0$, for which the measurement, which we denote as ${\bf M}_0$, has Kraus operators
\begin{align}\label{eqn34}
\hat A_1\otimes \hat B_1&=I_A\otimes[0]_B,\nonumber\\
\hat A_2\otimes \hat B_2&=\sqrt{1-q\,}[0]_A\otimes[1]_B,\nonumber\\
\hat A_3\otimes \hat B_3&=\left(\sqrt{q\,}[0]_A+[1]_A\right)\otimes\ket{0}_B\bra{1}.
\end{align}
This can be implemented by the following LOCC protocol using two rounds of communication: Bob measures $\hat M_1$ (evaluated at $\epsilon=0$, of course), they terminate the protocol with his first outcome, but upon getting his second outcome, Alice measures with $\hat M_2$. They terminate upon her first outcome, but if she gets her second outcome, Bob does a spin-flip on his qubit, and then they are done. If, on the other hand, they are just trying to implement the $\epsilon\rightarrow0$ limit of the measurement having Kraus operators equal to the positive semi-definite square roots of the POVM elements of \eqref{eqn32}---these Kraus operators are all diagonal in the $\ket{0},\ket{1}$ basis---then Bob has no need for that final spin-flip and this measurement can be done with only one round of communication. Thus, we see that for $0<\epsilon<1$, there exists a measurement that can only be done exactly with LOCC by using an infinite number of rounds, but that can be approximated within $O(\epsilon)$ by an LOCC protocol using only one round of communication. Indeed, using the distance measure of \eqref{eqn05}, we calculate explicitly that $d({\bf M}_\epsilon,{\bf M}_0)\le2\epsilon(1-q)/(1-q\epsilon)$.

In the next section, we demonstrate, perhaps unsurprisingly, that the example just discussed is not an isolated case.

\section{Generalizing the example of Section~\ref{sec3}}\label{sec4}
Consider the following sequence of measurements on any bipartite system that Alice and Bob will take turns performing. For $j=1,\cdots,L$, define
\begin{align}\label{eqn41}
M_{2j-1}&=\left\{B_{2j-1},B_{2j}\right\},\nonumber\\
M_{2j}&=\left\{A_{2j},A_{2j+1}\right\},
\end{align}
where Bob starts with $M_1$, Alice follows with $M_2$ upon Bob's second outcome, and then the parties sequentially alternate turns. As in the preceding section, the first outcome of each measurement ($B_{2j-1}$ or $A_{2j}$) terminates the protocol, whereas the second outcome ($B_{2j}$ or $A_{2j+1}$) is always followed by the other party measuring next. Finally, let
\begin{align}\label{eqn42}
A_1&=I_A,\notag\\
B_{2L}&=\sqrt{\epsilon}\hat B_{2L-2}^{-1},\nonumber\\
A_{2L+1}&=\sqrt{q}\hat A_{2L-1}^{-1}.
\end{align}
where the cumulative action of each party starting from their first measurement is given by
\begin{align}\label{eqn43}
\hat B_{1}&=B_1,\notag\\
\hat A_{2}&=A_2,\notag\\
\hat B_{i}&=B_iB_{2k_1}\cdots B_{4}B_{2}B_0,~2\le i\le2L,\notag\\
\hat A_{i}&=A_iA_{2k_2+1}\cdots A_{5}A_{3}A_1,~3\le i\le2L+1.
\end{align}
and we have set $B_0=I_B$. Here, $k_1$ is the largest integer such that $2k_1<i$, and $k_2$ is the largest integer such that $2k_2+1<i$.

\begin{figure}
\includegraphics{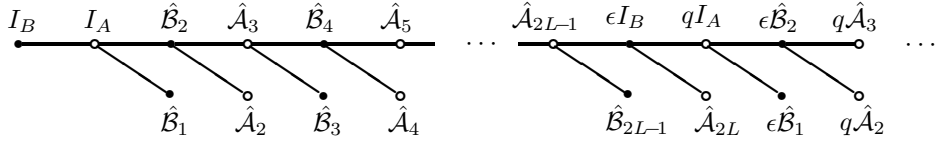}
\caption{\label{fig0}LOCC tree depicting an infinite-round protocol involving the individual local measurements of \eqref{eqn41} and implementing the overall measurement ${\bf M}_\epsilon$ of \eqref{eqn45}. Each node is labeled by the POVM element corresponding to that party's cumulative action to that point in the protocol, and each leaf can be followed by an isometry for each party to adjust Kraus operators, if desired. The sequence of individual measurements repeats itself after each node having label proportional to $I_A$.}
\end{figure}

We require these operators to satisfy the following constraints:
\begin{enumerate}
  \item\label{itm1}Each $A_{2j+1}$ and $B_{2j}$ is full-rank, so that the inverses in \eqref{eqn42} exist.
  \item \label{itm2}Comparing \eqref{eqn42}, we must choose $\epsilon$ ($q$) small enough that $B_{2L}$ ($A_{2L+1}$) can be part of a valid measurement. This means $\epsilon<\epsilon^\ast:=\lambda_B^{-2}$ $\left(q<q^\ast:=\lambda_A^{-2}\right)$, where $\lambda_{B}$ ($\lambda_{A}$) is the largest singular value of $\hat B_{2L-2}^{-1}$ ($\hat A_{2L-1}^{-1}$).
  \item\label{itm3}Define $\hat \AC_i=\hat A_i^\dag\hat A_i$ and $\hat \BC_i=\hat B_i^\dag\hat B_i ~\forall{i}$. Then for $j=1,\cdots,L$ require
\begin{align}\label{eqn44}
\hat \AC_{2j-1}=\hat\AC_{2j}+\hat\AC_{2j+1},\notag\\
\hat \BC_{2j-2}=\hat\BC_{2j-1}+\hat\BC_{2j}.
\end{align}
The purpose of these conditions is to ensure that \eqref{eqn00} is always satisfied, which is required for any valid LOCC protocol. Note also that inserting the last two lines of \eqref{eqn42} into the last two lines of \eqref{eqn43}, respectively, we have 
\begin{align}\label{eqn49}
\hat\AC_{2L+1}=qI_A,\notag\\
\hat\BC_{2L}=\epsilon I_B.
\end{align}
  \item\label{itm4}For $\epsilon>0$, no two of the $\hat\AC_i$ are proportional, and no two of the $\hat\BC_i$ are proportional.
\end{enumerate}
Note that it is easy to satisfy all these constraints simultaneously. In fact, randomly choosing operators under the constraints of item~\ref{itm1} and \ref{itm3} will almost certainly also satisfy item~\ref{itm4}. The overall separable measurement is defined in terms of Kraus operators as
\begin{align}\label{eqn45}
{\bf M}_\epsilon=\left\{\frac{1}{\sqrt{1-q\epsilon}}\hat A_i\otimes\hat B_i\right\}_{i=1}^{2L}.
\end{align}
This measurement ${\bf M}_\epsilon$ satisfies the following conditions:
\begin{itemize}
   \item Due to item~\ref{itm4}, above, we have that $\Delta=0$. Hence, by Theorem~\ref{thm1}, measurement ${\bf M}_\epsilon$ cannot be implemented by any finite-round LOCC protocol.
  \item  ${\bf M}_\epsilon$ is implemented successfully by the infinite-round LOCC protocol consisting of the sequence of local measurements $M_j$, for which the first outcome is always taken to terminate the protocol and the second outcome is followed by the other party measuring with $M_{j+1}$. Bob starts with $M_1$ and the sequence of measurements continues without end. By \eqref{eqn49}, this sequence effectively starts over at $q\epsilon I_A\otimes I_B$ with Alice's second outcome of $M_{2L}$. Therefore, Bob can just start again after this outcome by performing $M_1$, and the parties continually cycle through these $2L$ measurements. In the limit of infinite rounds, they will have successfully implemented ${\bf M}_\epsilon$, just as we saw for the $L=2$ example in the preceding section. See Figure~\ref{fig0} for a depiction of the corresponding LOCC tree.
  \item In the limit $\epsilon\rightarrow0,~\hat \BC_{2L-2}=\hat\BC_{2L-1}$ by \eqref{eqn44}, and we recover a protocol of $2L-2$ rounds ($2L-3$ rounds if Kraus operators $\hat B_{2L-1}=\hat B_{2L-2}$ in this limit) implementing a measurement which we denote as ${\bf M}_0$. By continuity of the distance measure of \eqref{eqn05}, we have that ${\bf M}_\epsilon$ continuously approaches ${\bf M}_0$ as $\epsilon$ approaches zero. Hence, for small $\epsilon$ the original measurement, which requires infinite rounds for LOCC, can be closely approximated by this finite-round protocol.
\end{itemize}
We have thus constructed an infinite class $\CC$ of measurements of the form ${\bf M}_\epsilon$ given in \eqref{eqn45}, where each element in this class is such that it cannot be implemented by finite-round LOCC, but can be implemented using an infinite number of rounds. For any finite $L$, there exists a sub-class $\CC_L\subset\CC$ for which the infinite-round LOCC protocol that implements one of the measurements in $\CC_L$ consists of a continually repeating sequence of $2L$ local measurements (of course, the limit as $L$ goes to infinity can also be used to construct valid measurements in $\CC$). In addition, for small $\epsilon<\epsilon^\ast$, these measurements ${\bf M}_\epsilon$ can be closely approximated by a measurement of either $2L-2$ or $2L-3$ rounds. An explicit example from this class of measurements with $L=2$ was given in Section~\ref{sec3}.

We now move to the next section, where we use what we've just learned to show that every finite-round LOCC is the limit of a sequence of separable measurements, where each measurement in that sequence can be implemented by LOCC, but only with the use of an infinite number of rounds.

\section{Every finite-round LOCC is arbitrarily close to separable measurements that require an infinite number of rounds}\label{sec5}
Given an arbitrary measurement ${\bf M}_0\in$ LOCC$_{\mathbb{N}}$, we wish to construct a continuous sequence of measurements, ${\bf M}_\epsilon$, such that
\begin{enumerate}[label=\alph*)]
  \item\label{itmA} $\lim_{\epsilon\rightarrow0}{\bf M}_\epsilon={\bf M}_0$; and
  \item\label{itmB} $\exists\epsilon^\ast>0$ such that ${\bf M}_\epsilon\in$ LOCC$_\infty$ for all $0<\epsilon<\epsilon^\ast$.
\end{enumerate}
We will do so by considering an LOCC protocol implementing ${\bf M}_0$ of that protocol a continually repeating sequence of $2L$ measurements of the kind described in the preceding section (the choice of $L$ can differ from one leaf to the next, and can even be chosen to be infinite, such choices do nothing to alter our conclusions). In order to be sure that item~\ref{itmB} is satisfied, we require that all leaf nodes  in the resulting infinite tree, with their respective POVM elements $\hat\AC_{jm}^{(\epsilon)}\otimes\hat\BC_{jm}^{(\epsilon)}$, are such that no two $\hat\AC_{jm}^{(\epsilon)}$ are proportional, and no two $\hat\BC_{jm}^{(\epsilon)}$ are proportional (that is, for example, $\forall{i,j,m,n}~\hat\AC_{jm}^{(\epsilon)}\not\propto\hat\AC_{in}^{(\epsilon)}$). This will imply that $\Delta=0$ for this measurement ${\bf M}_\epsilon$, and then by Theorem~\ref{thm1}, ${\bf M}_\epsilon\in$ LOCC$_\infty$.

Before appending these repeating sequences, there are two types of modifications we will make to the original tree. The first will be to alter those leaf nodes that are repeated---that is, which correspond to the same local POVM element---so that the modified tree will have no repeated leafs: after these modifications, every $A$ leaf will be distinct from every other one, and the same will hold for the $B$ leafs. The second type of modification has to do with the following issue: suppose there is a node in the original tree for which the cumulative action by Alice is given by Kraus operator $\hat A_j=\ket{\psi_j}\bra{\phi_j}$; that is, this operator is rank-$1$. The corresponding POVM element associated with this node is $\hat\AC_j=\eta_j\ket{\phi_j}\bra{\phi_j},~\eta_j=\inpd{\psi_j}{\psi_j}$, and this cannot be changed (apart from scaling factors) by any subsequent measurements that Alice performs. Since our strategy will be to create new and distinct POVM elements by appending measurements subsequent to the leafs of the original tree, starting with a rank-$1$ operator will be problematic. Therefore, we will alter each rank-$1$ outcome to increase its rank. All these modifications will be done in a way such that in the limit as $\epsilon\rightarrow0$, the original tree is recovered. 

Since a rank-$1$ $\hat\AC_j$ is unchanged by subsequent measurements, we may assume without loss of generality that Alice makes no further measurements after such an outcome in the original protocol. It may subsequently be necessary, however, for Alice to perform isometries to alter individual Kraus operators. Hence, each of Alice's rank-$1$ outcomes may be followed by a measurement by Bob, and for each outcome of the latter measurement, an isometry (or a measurement for which each outcome is proportional to an isometry) by Alice. Note that under these circumstances, Alice's isometries will terminate the protocol, so will be located at leaf nodes, and these leafs will be repeated, all corresponding to the same rank-$1$ POVM element $\hat \AC_j$. Therefore, these leafs will each be modified as indicated in the preceding paragraph.

Our strategy will be the same for both types of modifications. Every one of Alice's nodes that is to be modified will be modified as
\begin{align}\label{eqn50}
\hat\AC_j\rightarrow \hat\AC_j^{(\epsilon)}=\left[(1-\epsilon)\hat\AC_j+\epsilon\rho_j\right],~\rho_j\ge0, ~\rho_j\not\propto\hat\AC_j,
\end{align}
and then, one of that node's siblings is also modified to
\begin{align}\label{eqn55}
\hat\AC_i\rightarrow \hat\AC_i^{(\epsilon)}=\left[(1-\epsilon)\hat\AC_i+\epsilon\rho_i\right],~\rho_i\ge0, ~\rho_i\not\propto\hat\AC_i.
\end{align}
These modifications maintain \eqref{eqn00} as long as $\rho_i+\rho_j=\hat\AC_i+\hat\AC_j$, which we shall therefore require. They correspond to a modification of Kraus operators by $\hat A_j=U_j\sqrt{\hat\AC_j}\rightarrow U_j\sqrt{(1-\epsilon)\hat\AC_j+\epsilon\rho_j}$ and $\hat A_i\rightarrow U_i\sqrt{(1-\epsilon)\hat\AC_i+\epsilon\rho_i}$. Note that there is no problem with modifying a given node more than once, so we need not worry about running out of siblings when more than one sibling needs to be modified. That $\rho_j\not\propto\hat\AC_j$ guarantees $\hat\AC_j^{(\epsilon)}$ has rank exceeding $1$, as required. It also guarantees that pairs $\hat\AC_j,\rho_j$ (or $\hat\AC_i,\rho_i$) are rays defining a plane, and this observation will inform how the $\rho_j$ are chosen, see below.

If any of these modified $A$-nodes, say $\hat\AC_i^{(\epsilon)}$, is not at the location of a leaf in the original tree, it may be followed by subsequent $A$-measurements, so these descendant $A$ nodes must be altered so as to preserve \eqref{eqn00}. For the nearest descendent $A$ measurement, replace all but one of the outcomes as $\hat\AC_k\rightarrow(1-\epsilon)\hat\AC_k$ (and then multiply by $1-\epsilon$ all $A$ nodes descendant from these), and then for that one remaining outcome, say $l$, replace $\hat\AC_l\rightarrow(1-\epsilon)\hat\AC_l+\epsilon\rho_i$. Then repeat this process for the next descendant $A$ measurement from that $\hat\AC_l$ node, and continue this process until all descendant $A$ nodes are modified appropriately. This procedure ensures \eqref{eqn00} is maintained for all $A$ descendants of $\hat\AC_i^{(\epsilon)}$, ensuring that the tree continues to represent a valid LOCC protocol, and that the original protocol is recovered for $\epsilon\rightarrow0$. Of course, Bob's nodes are treated in a way completely analogous to what we have just described for Alice's.

We will choose the $\rho_j$ such that all leaf nodes are distinct (our method of doing so is discussed below, in the last paragraph of this section). Multiple leaf nodes share the same parent, however, so this modified protocol has $\Delta\ge1$, as it must given that it is still a finite-round protocol. The next step is to append, after each leaf node, an infinite branch corresponding to repeating sequences of measurements like those described in Section~\ref{sec4}. What results is an infinite-round LOCC protocol implementing a separable measurement corresponding to POVM elements $\hat\AC_{jm}^{(\epsilon)}\otimes \hat\BC_{jm}^{(\epsilon)}$, where we have denoted the POVM elements appearing along the branch appended to leaf $j$ by $\hat\AC_{jm}^{(\epsilon)},\hat\BC_{jm}^{(\epsilon)}$. This separable measurement will be an element of LOCC$_\infty$ if no two of the $\hat\AC_{jm}^{(\epsilon)}$ are proportional and no two of the $\hat\BC_{jm}^{(\epsilon)}$ are proportional. Let us now see how this can be accomplished.

\begin{figure}
\includegraphics{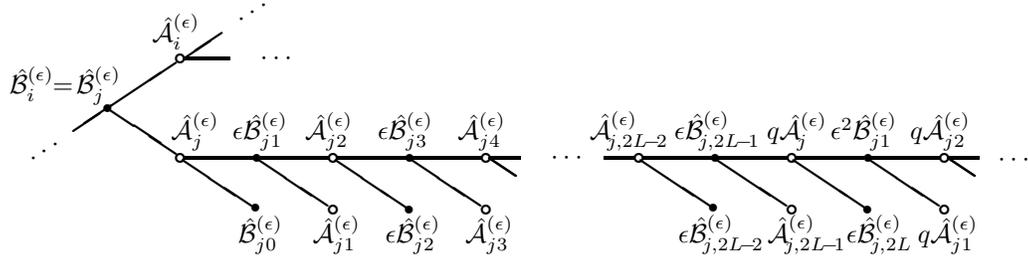}
\caption{\label{fig1}The infinite branch that is appended to leaf node $\hat\AC_j$ of the original tree, which itself was previously modified to $\hat\AC_j^{(\epsilon)}$, as described in the text. All other leaf nodes of the original tree, including the $\hat\AC_i^{(\epsilon)}$ node shown in the figure, have similar infinite branches appended to them in the final, infinite tree.}
\end{figure}

For each of Alice's (Bob's) leaf nodes, choose even integer $2L$, the length of the repeating sequence to be appended to that leaf. This repeating sequence will consist of Bob and Alice alternating two-outcome measurements, where the first outcome of each measurement terminates the protocol, and the second outcome is followed by the other party's next measurement. We require that the measurement operator for that second outcome does not decrease the rank of the cumulative Kraus operator, a requirement that is necessary if the sequence is to repeat itself. This requirement is essentially equivalent to item \ref{itm1} above \eqref{eqn44}, along with the recognition that in general, the root of this branch (the node to which this branch is appended) will not itself be full rank (when it is not full rank, the inverse in \eqref{eqn42} is to be replaced by the pseudo-inverse, sometimes referred to as the ``inverse on the support" of the given operator). Note that item~\ref{itm2} above \eqref{eqn44} constrains $\epsilon<\epsilon^\ast$, but since $\epsilon^\ast>0$ there will always be a non-vanishing range for $\epsilon$. We also require that for the first measurement after the original leaf to which this branch is being appended, that second outcome has POVM element proportional to $\epsilon$. This ensures that in the limit $\epsilon\rightarrow0$, the appended branch effectively truncates at that original leaf, recovering the original tree in this limit. This will be clear when one notes the position of the node $\epsilon\hat\BC_{j1}^{(\epsilon)}$ in Figure~\ref{fig1} and realizes that by \eqref{eqn00}, $\lim_{\epsilon\rightarrow0}\hat\BC_{j0}^{(\epsilon)}=\lim_{\epsilon\rightarrow0}\hat\BC_j^{(\epsilon)}=\hat\BC_j$.

It only remains to show that the appended measurements can be chosen so as to yield a protocol implementing an overall measurement having $\Delta=0$, so as to ensure that this measurement is in LOCC$_\infty$. Consider the repeating sequence of measurements shown in Figure~\ref{fig1}. The labels indicate the cumulative POVM elements associated with each node (operators $\hat\AC_j^{(\epsilon)},\hat\BC_j^{(\epsilon)}$ may each be equal to their respective operators in the original tree, $\hat\AC_j,\hat\BC_j$, or they may have been modified to $O(\epsilon)$ as previously described). These operators satisfy a modified version of \eqref{eqn44}, which for $m=1,\cdots,L$ reads
\begin{align}\label{eqn51}
\hat \AC_{j,2m-2}^{(\epsilon)}=\hat\AC_{j,2m-1}^{(\epsilon)}+\hat\AC_{j,2m}^{(\epsilon)},\notag\\
\hat \BC_{j,2m-1}^{(\epsilon)}=\hat\BC_{j,2m}^{(\epsilon)}+\hat\BC_{j,2m+1}^{(\epsilon)},
\end{align}
with $\hat \AC_{j0}^{(\epsilon)}=\hat \AC_{j}^{(\epsilon)}$, and \eqref{eqn49} is modified to
\begin{align}\label{eqn52}
\hat\AC_{j,2L}^{(\epsilon)}=q\hat\AC_{j}^{(\epsilon)},\notag\\
\hat\BC_{j,2L+1}^{(\epsilon)}=\epsilon\hat\BC_{j1}^{(\epsilon)}.
\end{align}

Recall that $\hat\AC_j^{(\epsilon)}=(1-\epsilon)\hat\AC_j+\epsilon\rho_j$. As $\epsilon$ varies, the ray generated by $\hat\AC_j^{(\epsilon)}$ traces out a two-dimensional cone between rays $\AC_j$ and $\rho_j$, and let us then choose all operators $\hat\AC_{jm}^{(\epsilon)}$ for this fixed $j$ to lie in the same plane as this cone. For example, we could set $q=2^{-L}$ and then for $m=1,\cdots,L-1$,
\begin{align}\label{eqn54}
\hat\AC_{j,2m}^{(\epsilon)}&=2^{-m}\hat\AC_j+\frac{1}{2}\left(\frac{2}{3}\right)^m\!\epsilon\rho_j,\notag\\
\hat\AC_{j,2m+1}^{(\epsilon)}&=2^{-m}\hat\AC_j+\left(\frac{2}{3}\right)^m\!\epsilon\rho_j,\notag\\
\end{align}
and
\begin{align}\label{eqn55}
\hat\AC_{j,2L}^{(\epsilon)}&=2^{-L}\hat\AC_j+\left(\frac{2}{3}\right)^{L-1}\!\epsilon\rho_j,
\end{align}
which together satisfy the $A$ parts of \eqref{eqn51} and \eqref{eqn52}. Similar choices can be made for the $B$ operators, as well.

Therefore, we have that for each specific leaf in the original tree, $\hat\AC_{j}$ or $\hat\BC_{j}$, all of a given party's POVM elements descended from that leaf in the extended, infinite tree lie in a given plane in the space of that party's operators, and that plane stays fixed as $\epsilon$ varies. Each of these planes lies in a space of dimension at least four. Therefore it is easy to choose the $\rho_j$ such that no two of these planes intersect with each other, unless they share the same operator from the original tree, $\hat\AC_j\propto\hat\AC_i$. In this case, the planes can be chosen (by choosing $\rho_i$ to not lie in the plane connecting rays $\hat\AC_j$ and $\rho_j$) so that the only intersection is precisely at that single ray, $\hat\AC_j$. In addition when two (or more) of these planes do intersect, that intersection, $\hat\AC_j$, is never a leaf or parent to a leaf in this final tree, since it was a repeated leaf in the original tree so has been modified to $\hat\AC_{j}^{(\epsilon)}\ne\hat\AC_j$ and $\hat\AC_{i}^{(\epsilon)}\ne\hat\AC_i$, where $\hat\AC_{i}^{(\epsilon)}\ne\hat\AC_j^{(\epsilon)}$. That is, this intersection never contains a POVM element of the overall measurement corresponding to the final, infinite-round protocol, and thus every ray generated by these POVM elements has multiplicity equal to one for this measurement. Hence, for each $0<\epsilon<\epsilon^\ast$, we have constructed an LOCC protocol implementing a measurement ${\bf M}_\epsilon$ for which $\Delta=0$, implying that ${\bf M}_\epsilon\in$ LOCC$_\infty$. Furthermore, $\lim_{\epsilon\rightarrow0}{\bf M}_\epsilon={\bf M}_0$, the latter being the original, arbitrarily chosen, finite-round measurement. This proves our claim that every finite-round LOCC protocol implements a measurement that is the limit of a continuous sequence of measurements ${\bf M}_\epsilon\in$ LOCC$_\infty$. Theorem~\ref{thm4} then follows directly. (It may be easier to visualize all of this by taking a section along the surface in which all operators have unit trace. Then, $\hat\AC_j,\rho_j$ define a line along which all these descendant POVM elements lie. If the system is a qubit, then these lines lie within the three-dimensional Bloch sphere, and for any other system this space is of strictly higher dimension. It is then clearly easy to choose all these line segments such that they do not intersect, except possibly at those points $\hat\AC_j$ which are not POVM elements of the final measurement.)

\section{Conclusions}\label{sec6}
In summary, we have proved a necessary condition for finite-round LOCC, Theorem~\ref{thm1}, and then provided an example that demonstrates this condition can be violated for bipartite systems when using an infinite number of rounds. We then constructed a general class of infinite-round LOCC protocols acting on bipartite systems, each of which violates this necessary condition, and which therefore implement measurements that cannot be performed in any finite number of rounds of communication between the parties. These measurements can, of course, be approximated increasingly well by utilizing more and more rounds, and the protocols we've devised provide a simple way to see how the error incurred by finite-round approximations can be successively reduced by including additional rounds.

We then showed that every finite-round LOCC protocol is the limit of a continuous sequence of measurements, where each measurement in that sequence can be implemented by LOCC, but only with the use of an infinite number of rounds. This includes so-called ``zero-round" protocols, in which the parties both measure but do not communicate, demonstrating the existence of measurements that can only be implemented by LOCC by using an infinite number of rounds, but that can nonetheless be well approximated without using any communication at all. These results constitute a constructive proof of Theorem~\ref{thm4}.

One of the implications of Theorem~\ref{thm4}, noted in Corollary~\ref{cor1}, is that LOCC$_{\mathbb{N}}$ has empty interior. In apparent contradiction, it was shown in \cite{WinterLeung} that LOCC$_{\mathbb{N}}$ has a non-empty interior. These results are, however, not in contradiction, the reason being that our definitions of LOCC$_{\mathbb{N}}$ are not the same: whereas we consider quantum \textit{measurements}, they consider quantum \textit{channels}. While we have shown here that every measurement ${\bf M}_0\in$ LOCC$_{\mathbb{N}}$ has measurements ${\bf M}_\epsilon\in$ LOCC$_\infty$ arbitrarily close to it, it may be (indeed, must be, for the example given in \cite{WinterLeung}) that each of those nearby ${\bf M}_\epsilon$ corresponds to a quantum channel that has a different representation in terms of product Kraus operators, where that different representation is a measurement that \textit{can} be implemented by finite-round LOCC. Therefore, these nearby measurements, ${\bf M}_\epsilon\in$ LOCC$_\infty$, correspond to quantum channels $\EC_\epsilon\in$ LOCC$_{\mathbb{N}}$. This explains how the set of quantum channels implementable by finite-round LOCC can have non-empty interior, even when the interior of the set of quantum measurements that can be so implemented is empty.

Whereas we have shown in the appendix that the bound in Theorem~\ref{thm1} is tight for bipartite systems, we have recently been able to show that it is \textit{not} tight for $P=3$ (and presumably also for $P>3$). There are, however, many interesting, related questions that remain to be explored for systems having more than two parties, so we leave discussion of these results to a future publication.

We have seen that the distinction between $\Delta=0$ and $\Delta\ge1$ is significant in that it provides a separation between LOCC$_\mathbb{N}$ and LOCC$_\infty$, where $\Delta$ is defined in \eqref{eqn21}. One may ask if a more detailed analysis using this approach of counting distinct rays can provide additional information in the case $\Delta\ge1$, perhaps concerning how many rounds are required to implement a given measurement by LOCC. It appears that by itself, this may only yield limited results, though it may be more fruitful when combined with the notion of counting distinct extreme rays, as was done in the context of a different necessary condition for LOCC \cite{myExtViolate1,myUSD}. We can, however, show that when $\Delta=1$, at least two rounds of communication are necessary unless $N\le3$, but that there exist $\Delta=1$ measurements with any $N\ge4$, including $N\rightarrow\infty$, for which two rounds is sufficient (a protocol similar to that given for $N\rightarrow\infty$ in the appendix is one way to construct such examples, and one can there replace the initial infinite-outcome measurement by one with only a finite number of outcomes).

It is not difficult to see that $\textrm{LOCC}_\mathbb{N}$ is a convex set, and the notion of convexity points to several interesting issues. Even though the foregoing results indicate that $\textrm{LOCC}_\mathbb{N}$ and $\textrm{LOCC}_\infty$ are intimately intertwined with each other, it is easy to construct examples of pairs of measurements in LOCC$_\infty$ for which there is a continuous range of convex combinations of that pair which lie in LOCC$_\mathbb{N}$. That is, there are pairs of measurements ${\bf M}_1,{\bf M}_2\in$ LOCC$_\infty$ and parameters $0<\lambda_-<\lambda_+<1$ such that for $\lambda_-\le\lambda\le\lambda_+$, the measurement ${\bf M}=\sqrt{\lambda}{\bf M}_1+\sqrt{1-\lambda}{\bf M}_2\in$ LOCC$_\mathbb{N}$ (the square roots appear here because it is the POVMs that obey convexity, rather than measurements as we have here defined them in terms of Kraus operators, and if a POVM element is multiplied by $\lambda$, this is equivalent to multiplying the corresponding Kraus operator by $\sqrt{\lambda}$). At the same time, there exist measurements in LOCC$_\mathbb{N}$ that cannot be written as a convex combination of measurements in LOCC$_\infty$ at all. It therefore appears that there are many interesting questions left to be understood about the detailed structure of LOCC.

Finally, we note that the measurement protocol of \cite{FortescueLo} for implementing a random distillation of the tripartite $W$ state to a maximally entangled state on a randomly chosen pair of the original three parties is very similar to the protocols we have described in Section~\ref{sec4}, for the case $L\rightarrow\infty$. Are there also examples of measurements in LOCC$_\infty$ of the form constructed in Section~\ref{sec5} that arise out of physically motivated tasks, such as random distillation? It would certainly be of interest to find such examples, which would then correspond to physical tasks that cannot be exactly implemented by any finite-round protocol but which can be approximated closely in some relatively small number of rounds.

\noindent\textit{Acknowledgments} --- The author would like to thank Bob Griffiths, Maris Ozols, and Dan Stahlke for very helpful comments. This work has been supported in part by the National Science Foundation through Grant No. 1205931.
\appendix*
\section{Achieving $\Delta=1$ for bipartite systems}

We now show that for finite-round LOCC, $\Delta=1$ is achievable for $P=2$ and any number of distinct operators $\hat\KC_j$. If this number, $N$, is finite, consider a bipartite LOCC protocol where the parties alternate their two-outcome measurements. The first outcome of each of these measurements is terminal, the second outcome is followed by the other party measuring with a subsequent two-outcome measurement. Then, after $N-2$ such measurements, one party does a final two-outcome measurement, giving a total number of $N$ leaves in the full tree. It is not difficult to choose the $\hat\KC_j^{(\alpha)}$ labeling each leaf such that the corresponding ray $\lambda\hat\KC_j^{(\alpha)}$ is distinct from all the others, for each party $\alpha$, and for these leaf nodes to be consistent with a valid LOCC protocol. Each of these leaf nodes, say $\hat\KC_j^{(\alpha)}$, has a parent node $\hat\KC_j^{(\beta)}$ as its counterpart, with $\beta\ne\alpha$. One can also choose these parent nodes to be distinct from each other and from all that party's leafs, for each party, so that the measurement implemented by this protocol is such that $\forall{\alpha},~\hat\KC_j^{(\alpha)}\not\propto\hat\KC_i^{(\alpha)},~\forall{i\ne j}$, except of course for that final two-outcome measurement, for which that parent node has multiplicity equal to two. In this way, we have an LOCC protocol of $N-1$ rounds where every ray but one has multiplicity equal to unity, and thus $\Delta=1$, saturating the lower bound in Theorem~\ref{thm1} for every finite $N$.

When there are an infinite number of distinct operators, the lower bound can still be saturated. One way to do this is for the first party to start with an infinite outcome measurement, followed for each of her (distinct, by design) outcomes $\hat\KC_m^{(1)}=w_{m1}\hat\KC_1^{(1)}+w_{m2}\hat\KC_2^{(1)}$, by the second party doing a two-outcome measurement. For each of the latter measurements, one outcome terminates the protocol, and the other outcome is proportional to $\hat\KC_1^{(2)}=\hat\KC_2^{(2)}$, and is followed by the first party again measuring, this time with the two outcomes being $w_{m1}\hat\KC_1^{(1)}$ and $w_{m2}\hat\KC_2^{(1)}$. We see that all of the first party's operators are distinct---even though $\hat\KC_1^{(1)}$ and $\hat\KC_2^{(1)}$ each appear at an infinite number of leafs, they only appear as $\hat\KC_1^{(1)}\otimes\hat\KC_1^{(2)}$ and $\hat\KC_2^{(1)}\otimes\hat\KC_1^{(2)}$ so still have unit multiplicity. One can easily choose the rest of the second party's operators such that the ray $\hat\KC_1^{(2)}$ is the only one with multiplicity greater than one. Therefore, $\Delta=1$, as desired. Actually, this construction also works for saturating the bound when $N$ is finite, one just has to replace that initial measurement by one with a finite number of outcomes. These protocols utilize two rounds of communication, one for party $1$ to send her outcome $m$ to the other party, and the second round is to inform that first party whether or not to measure again.

%
%

\end{document}